%% file: anonymous-submission-latex-2026.tex
\documentclass[letterpaper]{article} 
\usepackage{aaai2026}  
\usepackage{times}  
\usepackage{helvet}  
\usepackage{courier}  
\usepackage[hyphens]{url}  
\usepackage{graphicx} 
\usepackage{natbib}  
\usepackage{caption} 
\usepackage{algorithm}
\usepackage{algorithmic}
\usepackage{amsmath,amssymb}
\usepackage{multirow}
\usepackage{booktabs}
\usepackage{xcolor}
\newcommand{\techName}{\emph{StrokeFusion}}

\usepackage{newfloat}
\usepackage{listings}
\DeclareCaptionStyle{ruled}{labelfont=normalfont,labelsep=colon,strut=off} 
\floatstyle{ruled}
\newfloat{listing}{tb}{lst}{}
\floatname{listing}{Listing}
\title{StrokeFusion: Vector Sketch Generation via \\ Joint Stroke-UDF Encoding and Latent Sequence Diffusion}
\author{
    Jin Zhou,
    Yi Zhou,
    Hongliang Yang,
    Pengfei Xu\thanks{Corresponding author, e-mail: xupengfei.cg@gmail.com},
    Hui Huang
}
\affiliations{
    College of Computer Science and Software Engineering, Shenzhen University, China 
}

\begin{document}

\maketitle




\input{Text/W_camera_ready}


\bibliography{aaai2026}

\input{Text/X_suppl}
\end{document}

%% file: Text/W_camera_ready.tex
\begin{abstract}
In the field of sketch generation, raster-format-trained models often produce non-stroke artifacts, while vector-format-trained models typically lack a holistic understanding of sketches, resulting in compromised recognizability. Moreover, existing methods struggle to extract common features from similar elements (e.g., animal eyes) that appear at varying positions across sketches. To address these challenges, we propose \techName, a two-stage framework for vector sketch generation. It contains a dual-modal sketch feature learning network that maps strokes into a high-quality latent space. This network decomposes sketches into normalized strokes and jointly encodes stroke sequences with Unsigned Distance Function (UDF) maps, representing sketches as sets of stroke feature vectors. Building upon this representation, our framework exploits a stroke-level latent diffusion model that simultaneously adjusts stroke position, scale, and trajectory during generation. This enables high-fidelity stroke generation while supporting stroke interpolation editing. Extensive experiments across multiple sketch datasets demonstrate that our framework outperforms state-of-the-art techniques, validating its effectiveness in preserving structural integrity and semantic features. Code and models will be made publicly available upon publication.
\end{abstract}

\section{Introduction}

Sketch generation, as an essential component of computational creativity, significantly accelerates concept visualization and rapid design iteration in numerous fields, including product design, animation, and interactive prototyping. Although humans effortlessly produce and interpret sketches by intuitively capturing holistic structures and local details, existing computational methods fall short of emulating this capability. These limitations manifest as difficulties in capturing global semantic structures, maintaining stroke-level control, and generating visually coherent sketches, thereby limiting their practicality in professional and creative workflows. Consequently, addressing these challenges through tailored computational paradigms is imperative.

Current sketch representations primarily exist in two formats: \textit{raster sketches} and \textit{vector sketches}.
Raster sketches render strokes directly as images and are learned through reconstruction of raster images~\cite{wang2024vq,ge2020creative,bhunia2022doodleformer}, allowing the use of mature image representation learning methods. However, these models often produce pixelation artifacts and non-stroke noise. More critically, they discard stroke trajectories, making it difficult to capture human drawing habits and natural stroke patterns, focusing solely on overall shape.

In contrast, vector sketches preserve clean and precise stroke trajectories that better simulate human drawing processes. Current vector-based approaches typically trace the entire stroke trajectory as polylines, recording pen-up and pen-down states. This leads to two modeling paradigms: modeling strokes using velocity-based relative coordinates~\cite{ha2017neural,das2023chirodiff}, or using absolute position representations~\cite{wang2023sketchknitter,bandyopadhyay2024sketchinr}. Relative coordinate modeling allows the model to learn common patterns more easily (e.g., eyes, wheels), since velocity vectors are often similar across these structures. However, these models must integrate velocity vectors to determine current positions, leading to error accumulation that makes it difficult to form closed loops or to determine position components accurately. Absolute coordinate modeling alleviates these issues but weakens the model's ability to capture stroke-wise commonalities, leading to inferior quality compared to velocity-based methods.

To address these challenges, we propose \techName, a two-stage framework that combines dual-modal feature encoding with latent space diffusion. In the first stage, we introduce \textit{stroke-UDF joint encoding}, which fuses vector primitives with rasterized unsigned distance fields (UDFs) to capture both geometric structure and stroke-level semantics. In the second stage, a \textit{stroke-level latent diffusion model} generates strokes in a non-autoregressive and order-invariant manner by jointly denoising position, scale, and trajectory in a structured latent space.

Our main contributions are:
\begin{itemize}
\item A dual-modal encoding scheme that integrates vector and raster features to enhance stroke representation while preserving spatial and semantic attributes.
\item A disentangled design that separates stroke layout prediction and shape synthesis, improving structural consistency and generation controllability.
\item A latent diffusion model that supports unordered, variable-length stroke generation, overcoming the limitations of sequential models.
\end{itemize}

Extensive experiments demonstrate that our method significantly outperforms baseline approaches by leveraging stroke-sketch hierarchical structures. Both quantitative metrics and qualitative comparisons reveal the advantages of our framework. Ablation studies on dual-modal encoding, stroke latent space, and spatial information disentanglement further validate the effectiveness of our design choices.

\begin{figure*}[t]
\centering
\includegraphics[width=\linewidth]{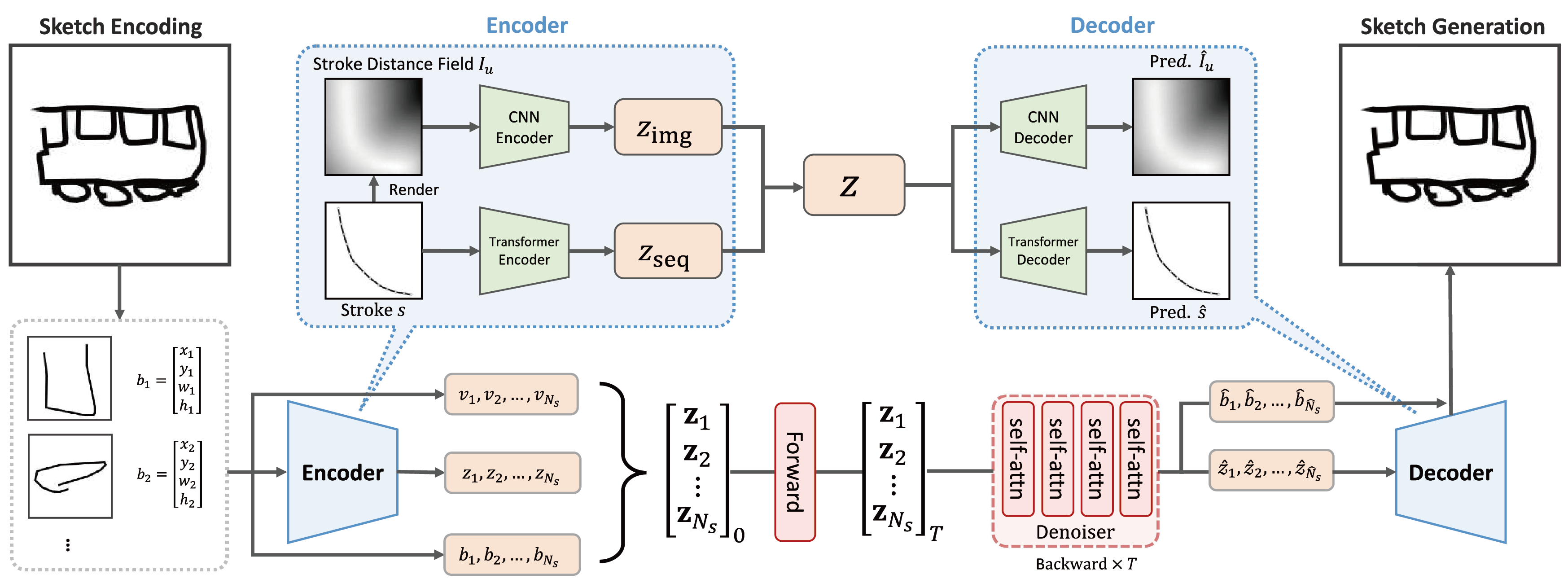}
\caption{The proposed \techName\ framework comprises two core components: 
1) \textbf{Dual-Modal Stroke Encoding}: Each stroke $s$ is processed through parallel encoding paths - a transformer-based sequence encoder handles geometric coordinates while a CNN processes the stroke distance field $I_n$. These modalities are fused into joint features $f$, trained via symmetric decoder networks that reconstruct both the original stroke ($s$) and distance field ($I_n$);
2) \textbf{Sketch Diffusion Generation}: All normalized strokes are encoded into latent vectors $z_i$, augmented with bounding box parameters $b^i = [x^i, y^i, w^i, h^i]$ and presence flags $v^i \in \{-1,1\}$. The diffusion model learns the distribution of stroke sequences $\{\mathbf{z}_1, ..., \mathbf{z}_N\}$ through $T$-step denoising training. During generation, the denoiser progressively refines noisy latents via reverse diffusion, with valid strokes ($v^i=1$) being decoded through inverse normalization of $\hat{b}^i$ to reconstruct the final sketch. The architecture maintains permutation invariance through order-agnostic sequence processing.}
\label{fig:pipeline}
\end{figure*}

\section{Related Work}

\subsection{Raster Sketch Generation}
Various methods focus on generating raster sketches. DoodlerGAN~\cite{ge2020creative} introduced a part-based GAN framework for learning the appearance of semantic parts in image space. DoodleFormer~\cite{bhunia2022doodleformer} improved structural understanding using a coarse-to-fine generation strategy with a two-stage Transformer. VQ-SGen~\cite{wang2024vq} implemented vector quantization to encode sketches into discrete stroke patches for autoregressive decoding. While these methods yield visually appealing results, they do not produce editable vector outputs. Our approach combines image-level supervision in shape modeling with the benefits of vector-format outputs.

\subsection{Vector Sketch Generation}
Ha et al.~\cite{ha2017neural} developed stroke-annotated polyline sequences with the Sketch-RNN model. Ribeiro et al.~\cite{ribeiro2020sketchformer}, Xu et al.~\cite{xu2024sketchformer++}, and Lin et al.~\cite{lin2020sketch} enhanced Transformer-based architectures for representation learning aimed at recognition and completion rather than generative modeling.
Qi et al.~\cite{qi2022generative} used Graph Convolutional Networks~\cite{kipf2016semi} for local structure modeling. Tiwari et al.~\cite{tiwari2024sketchgpt} tokenized sketches for GPT-style decoding, while Zang et al.~\cite{zang2023self} used Gaussian mixture latent spaces for structure-aware priors. Wang et al.~\cite{wang2023sketchknitter} represented vector sketches as offset sequences to improve diffusion efficiency. Das et al.~\cite{das2023chirodiff} approached stroke generation as continuous functions. However, many approaches overfit to superficial details, leading to inefficient training and poor generalization. Inspired by recent progress in 3D shape generation~\cite{zhang20233dshape2vecset}, our method encodes strokes into fixed-length feature vectors, treating them as unordered sets to handle variability in stroke count and order.

\subsection{Continuous Vector Sketch Generation}
Traditional methods often discretize sketches, overlooking the continuous nature of human drawing. To address this, several methods explore continuous representations. PMN~\cite{alaniz2022abstracting} used primitives from a graphical library, while B\'ezierSketch~\cite{das2020beziersketch} and Cloud2Curve~\cite{das2021cloud2curve} employed B\'ezier curves with adaptive orders. Stroke Clouds~\cite{ashcroft2023modelling} used diffusion models for third-order B\'ezier curves, but these techniques suffer from limited expressiveness.

Recent work focuses on fully continuous representations. SketchODE~\cite{das2022sketchode} applied Neural ODEs for stroke dynamics, and SketchINR~\cite{bandyopadhyay2024sketchinr} used implicit neural representations. However, these methods can miss holistic coherence. Our approach encodes entire strokes as learnable embeddings, enabling semantically coherent sketch composition while maintaining clean vector trajectories.

\subsection{Diffusion Models}
Our work is rooted in Diffusion Models~\cite{sohl2015deep}, which involve forward and reverse processes. Recent advances include continuous-space implementations~\cite{dhariwal2021diffusion,ho2020denoising} and discrete-space extensions like D3PM~\cite{austin2021structured}. We incorporate these advancements in diffusion modeling and align them with recent progress in 3D shape representation.

\section{Method}

\subsection{Problem Formulation and Representation}

Existing sketch generation approaches face a fundamental representation dilemma: point-sequence models capture stroke geometry but struggle with visual fidelity, while raster-based methods preserve appearance yet lose structure-level semantics. To bridge this gap, we propose a two-stage sketch generation framework comprising (1) \textit{stroke-UDF joint encoding}, which fuses vector stroke geometry and spatial structure, and (2) a \textit{stroke-level latent diffusion model} that synthesizes strokes in an unordered, non-autoregressive fashion. In this section, we detail our representation strategies for both vector and raster modalities.

\subsection{Vector Representation}
Depending on data sources, input strokes may be represented as B\'ezier curves or RDP-simplified polylines. To ensure representation consistency across different formats and abstraction levels, we resample each stroke into $N_p$ evenly spaced points along its trajectory. Thus, each stroke $s_j$ is represented as a sequence of points:
\begin{equation}\label{eq:stroke_def}
s_j = {p_1, p_2, \dots, p_{N_p}},\quad p_i = (x_i, y_i),
\end{equation}
where $(x_i, y_i)$ are normalized coordinates obtained with respect to the stroke's tight bounding square. This normalization guarantees scale invariance and facilitates stroke-wise feature extraction.

\subsection{Visual Representation}
To capture stroke shape in a raster-friendly manner, we convert each stroke into an unsigned distance field $I_u(g)$ using aggregated stroke density fields~\cite{alaniz2022abstracting}. For each segment defined by two consecutive points $x_i$ and $x_{i+1}$, we compute interpolated locations:
\begin{equation}\label{eq:interp_point}
p_i(r) = r x_i + (1 - r)x_{i+1}, \quad r \in [0,1].
\end{equation}
Then, the exponential decay of distance from any grid location $g$ to a stroke segment is defined as:
\begin{equation}\label{eq:segment_exp_dist}
d_i(g) = \max_{r \in [0,1]} \exp\left(-\gamma | g - p_i(r) |^2 \right),
\end{equation}
where $\gamma$ controls the sharpness of the field. Finally, the unsigned distance field $I_u(g)$ is defined as the maximal contribution over all segments:
\begin{equation}\label{eq:unsigned_distance_field}
I_u(g) = \max_{i \in {1,\dots,N_{p-1}}} d_i(g).
\end{equation}

\subsection{Stroke Embedding}

We adopt an encoder-decoder architecture to obtain neural embeddings for individual strokes. The dual-modal encoder jointly processes vector sequences and distance fields through two specialized branches, followed by feature fusion, as illustrated in Figure~\ref{fig:pipeline}.

\subsection{Preprocessing}
Our two-stage normalization process ensures geometric consistency:
\begin{itemize}
    \item \textbf{Sketch-level normalization}: center and scale the input sketches to $[-1, 1]$ while preserving their aspect ratios. This step standardizes the distribution of stroke bounding boxes.
    \item \textbf{Stroke-level normalization}: independently normalize each stroke's bounding box coordinates to $[0, 1]$ through linear scaling, maintaining individual aspect ratios for geometric relationship preservation.
    \item \textbf{Distance field rendering}: When rendering the unsigned distance field $I_u(g)$ using Equation~\ref{eq:unsigned_distance_field}, each stroke is additionally scaled by a factor of 0.8. This margin ensures that strokes do not get truncated by the canvas boundaries during rasterization.
\end{itemize}

\subsection{Vector Encoder} 
Given a stroke point sequence $s_j = {p_1, ..., p_{N_p}}$, each point $p_i = (x_i, y_i)$ is first projected via:
\begin{equation}
h_i^0 = \mathrm{Linear}(x_i, y_i) + \mathrm{PE}(i),
\end{equation}
where $\mathrm{Linear}(\cdot)$ maps coordinates to $\mathbb{R}^{d_h}$ and $\mathrm{PE}(\cdot)$ denotes sinusoidal positional encoding~\cite{vaswani2017attention}. These embeddings ${h_i^0}$ are processed through 6 stacked Transformer layers:
\begin{equation}
h_i^{l+1} = \mathrm{TransformerLayer}(h_i^l).
\end{equation}
The final output sequence ${h_i^6}$ is pooled via attention or average pooling to form a compact sequence-level feature $z_{\mathrm{seq}} \in \mathbb{R}^{d_{\mathrm{seq}}}$.

\subsection{Image Encoder} 
The unsigned distance field $I_u$ is encoded through 6 convolutional blocks with progressively increasing channel widths (from 4 to 128), each followed by ReLU activation:
\begin{equation}
z_{\mathrm{img}}^{(l+1)} = \mathrm{ReLU}(\mathrm{Conv2d}(z_{\mathrm{img}}^{(l)})).
\end{equation}
A global average pooling and linear projection compress the resulting feature map into a visual embedding $z_{\mathrm{img}} \in \mathbb{R}^{d_{\mathrm{img}}}$.

\subsection{Feature Fusion} 
The final fused representation $z_f$ combines geometry and visual semantics:
\begin{equation}
z_f = \mathrm{FC}(z_{\mathrm{seq}} \Vert z_{\mathrm{img}}),
\end{equation}
where $\Vert$ denotes concatenation along the feature dimension and $\mathrm{FC}(\cdot)$ is a fully connected layer projecting to $\mathbb{R}^{d_f}$.

\subsection{Stroke Decoders} 
Mirroring the encoder, the decoder comprises two parallel branches. The vector decoder expands $\hat{z}_p$ into per-point features using a feedforward network and applies 6 Transformer layers to reconstruct point sequences $\hat{s} = {\hat{p}_1, ..., \hat{p}_{N_p}}$, where $\hat{p}_i = (\hat{x}_i, \hat{y}_i, \hat{m}_i)$. In parallel, the image decoder upsamples $\hat{z}_u$ through 6 transposed convolutions to produce the reconstructed distance field $\hat{I}_u$. At inference time, only the vector decoder is used to generate strokes with actual drawable trajectories.

\subsection{Loss Functions}

\subsubsection{Vector-level Supervision.} We apply $L_2$ loss over coordinate pairs:
\begin{equation}\label{eq:vector_l2_loss}
\mathcal{L}_{\mathrm{vec}} = \frac{1}{N_p} \sum_{i=1}^{N_p} | (x_i, y_i) - (\hat{x}_i, \hat{y}_i) |_2.
\end{equation}

\subsubsection{Image-level Supervision.} 
The distance field output is optimized via:
\begin{equation}\label{eq:image_loss}
\mathcal{L}_{\mathrm{img}} = | I_u - \hat{I}_u |_2 + \mathcal{L}_{\mathrm{percep}}(I_u, \hat{I}_u),
\end{equation}
where $\mathcal{L}_{\mathrm{percep}}$ denotes the perceptual loss~\cite{zhang2018unreasonable}. A KL divergence regularization further aligns latent features:
\begin{equation}\label{eq:kl_loss}
\mathcal{L}_{\mathrm{KL}} = D_{\mathrm{KL}}\left(q(z_f|x),|,\mathcal{N}(0, I)\right).
\end{equation}

\subsubsection{Total Loss.} The combined training objective is:
\begin{equation}\label{eq:total_loss}
\mathcal{L}_{\mathrm{total}} = \lambda_{\mathrm{vec}}\mathcal{L}_{\mathrm{vec}}+ \lambda_{\mathrm{img}}\mathcal{L}_{\mathrm{img}} + \lambda_{\mathrm{KL}}\mathcal{L}_{\mathrm{KL}},
\end{equation}
where $\lambda_{\mathrm{vec}}$, $\lambda_{\mathrm{img}}$, and $\lambda_{\mathrm{KL}}$ are weighting hyperparameters. This multi-level supervision enables the model to learn both stroke geometry and perceptual plausibility, forming the basis for high-fidelity and editable sketch generation.

\subsection{Diffusion-Based Sketch Generation}\label{subsec:diffusion_model}

Building on the disentangled stroke representations learned in the previous stage, we now describe a generative model that synthesizes entire sketches from sets of latent strokes.

Our encoder maps each stroke into a compact latent space. Built on this, the Sketch Diffusion Generator models stroke collections and spatial relationships by generating unordered and variable-length sequences, as illustrated in Figure~\ref{fig:pipeline}.

\subsection{Training Process}
During training, each normalized stroke is encoded into a latent embedding $z_i$ using the dual-modal encoder. To model visibility and spatial configuration jointly, each embedding is augmented with its bounding box $b_i$ and a visibility flag $v_i \in \{-1, 1\}$, where $v_i = 1$ indicates a valid stroke and $v_i = -1$ denotes absence. The validity flag is introduced exclusively during the diffusion stage and is not supervised or predicted during encoder pretraining.

The bounding box $b_i = (x, y, w, h)$ includes the center position $(x, y)$ and width-height $(w, h)$ of the stroke, all normalized to $[0,1]$. During normalization and inverse normalization, scaling is performed isotropically using the larger of $w$ and $h$ to preserve the aspect ratio. The resulting composite latent vector is:
\begin{equation}
\mathbf{z}_i = [z_i,\, b_i,\, v_i].
\end{equation}
A sequence of such vectors $\{\mathbf{z}_1, ..., \mathbf{z}_{N_s}\}$ is passed into a diffusion model implemented as a 16-layer Transformer without positional encoding, which ensures permutation invariance. The model denoises these representations through $T$ steps in a Markov chain to learn the joint distribution over strokes and their spatial layout. The value $N_s$ is chosen as a conservative upper bound based on the 99th percentile of sketch stroke counts across the dataset; sketches exceeding this threshold are discarded during training.

\subsection{Generation Process}
\begin{figure}[t]
    \centering
    \includegraphics[width=\linewidth]{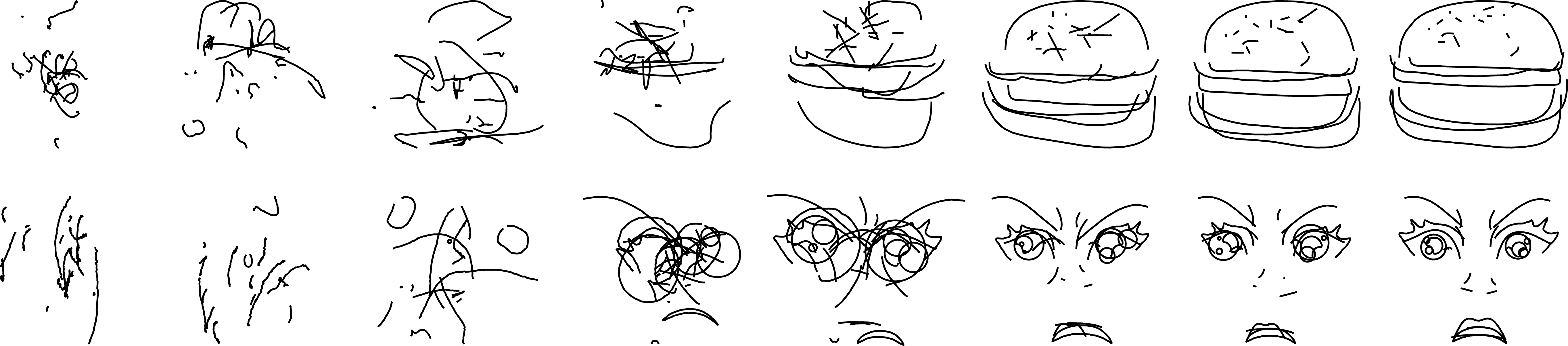}
    \caption{Two examples of the generation process in \techName. From left to right, the noise level progressively decreases. At each timestep, only strokes with presence confidence $\hat{v}_i > 0$ are visualized.}
    \label{fig:fig_diff_progress}
\end{figure}

\begin{figure}
\centering
\includegraphics[width=\linewidth]{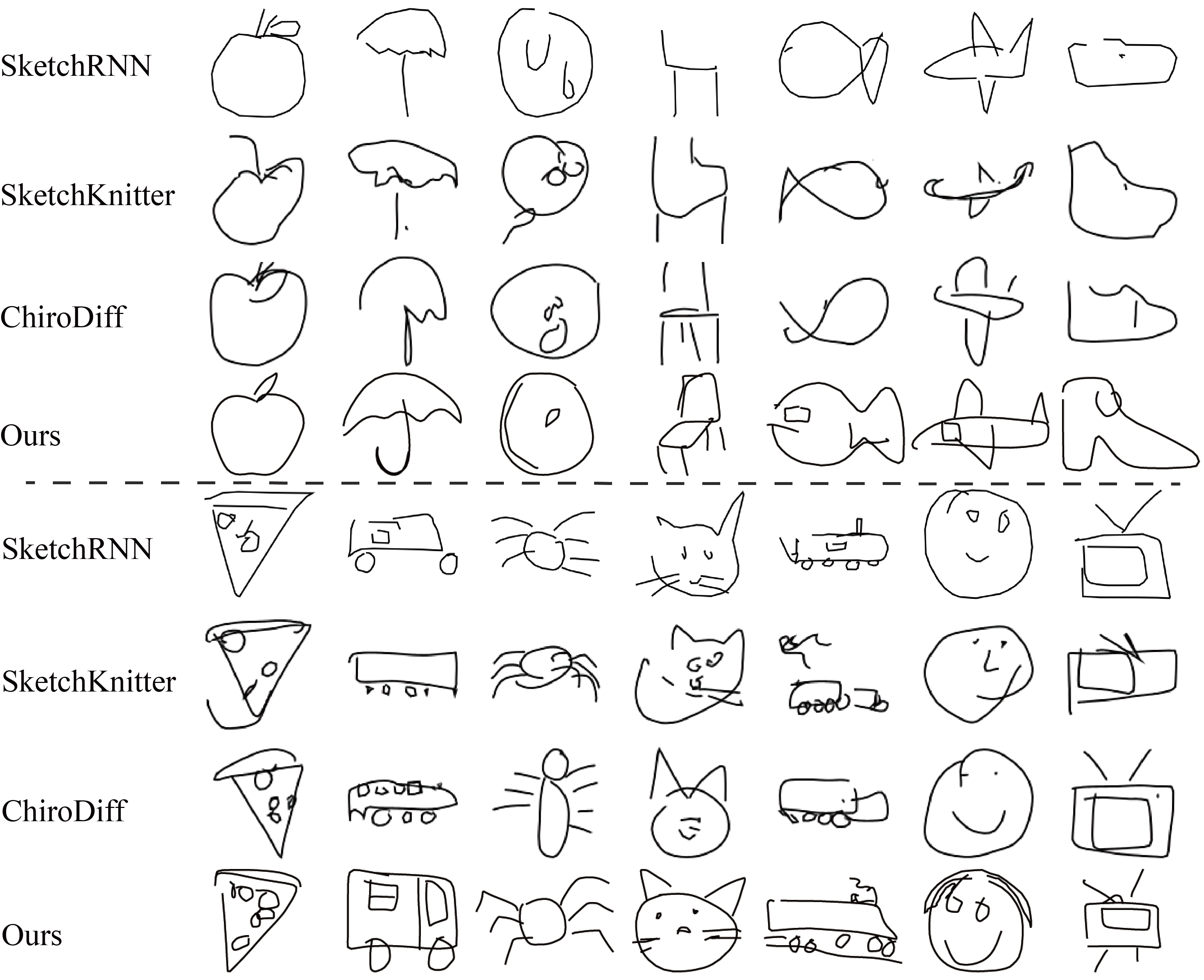}
\caption{Qualitative comparison of sketches generated by our method and the baselines across different categories in QuickDraw. Our method consistently produces more structurally coherent sketches with richer local details, particularly in complex, multi-stroke scenarios.}
\label{fig:exp-qualitative-quickdraw}
\end{figure}

At inference time, the diffusion model initializes from a noise sequence $\{\mathbf{z}_1, ..., \mathbf{z}_{N_s}\}_T$ and iteratively denoises it to obtain clean latent representations $\{\mathbf{z}_1, ..., \mathbf{z}_{N_s}\}_0$. Each denoised vector is processed by a shared MLP:
\begin{equation}
[\hat{z}_i,\, \hat{b}_i,\, \hat{v}_i] = \mathrm{MLP}_{\mathrm{split}}(\mathbf{z}_i), \label{eq:latent_split}
\end{equation}
where $\hat{v}_i$ is a continuous scalar. During decoding, a stroke is considered present if $\hat{v}_i > 0$, and discarded otherwise. Valid strokes are inversely normalized using $\hat{b}_i$ (with the same isotropic scaling rule) to recover global coordinates. These are then decoded into vector trajectories using the pretrained vector decoder. The resulting sketch is composed by overlaying all retained strokes in their correct positions.

Figure~\ref{fig:fig_diff_progress} illustrates two examples of the denoising process. As the noise level decreases, each stroke progressively refines its shape, spatial placement, and presence confidence by implicitly referencing other strokes within the composition. This self-organizing behavior emerges from the model’s joint reasoning over the entire stroke set.

Unlike existing vector sketch generation methods that treat all strokes as a concatenated trajectory~\cite{yang2021sketchgnn,das2023chirodiff,wang2023sketchknitter}, our design allows for a flexible number of strokes and diverse layout patterns. This is achieved by processing strokes as an unordered set without relying on positional encodings, thereby preserving permutation invariance throughout the generation process.

\subsection{Training Procedure}

We follow a two-stage training scheme: (1) pretrain the dual-modal encoder with reconstruction losses, and (2) train the diffusion model to synthesize coherent stroke layouts in latent space. The encoder is frozen in the second stage to ensure consistency.

\section{Experiments}
\subsection{Experimental Setup}


\begin{table*}[t]
\centering
\begin{tabular}{lccccccccc}
\toprule
\multirow{2}{*}[-0.5ex]{Method} & \multicolumn{3}{c}{$< 4$ strokes} & \multicolumn{3}{c}{$< 8$ strokes} & \multicolumn{3}{c}{$\geq 8$ strokes}\\
\cmidrule(lr){2-4}\cmidrule(lr){5-7}\cmidrule(lr){8-10}
 & FID$\downarrow$ & Prec$\uparrow$ & Rec$\uparrow$ & FID$\downarrow$ & Prec$\uparrow$ & Rec$\uparrow$ & FID$\downarrow$ & Prec$\uparrow$ & Rec$\uparrow$ \\
\midrule
SketchRNN & 31.61 & 0.49 & 0.45 & 36.98 & 0.58 & 0.44 & 40.67 & 0.55 & 0.40 \\
SketchKnitter & 23.17 & 0.52 & 0.48 & 27.07 & 0.57 & \underline{0.45} & 35.64 & 0.54 & 0.40 \\
ChiroDiff & \textbf{17.17} & \underline{0.61} & \underline{0.50} & \underline{23.84} & \underline{0.63} & \underline{0.45} & \underline{27.78} & \underline{0.62} & \underline{0.42} \\
\techName\ & \underline{19.53} & \textbf{0.71} & \textbf{0.58} & \textbf{18.99} & \textbf{0.69} & \textbf{0.61} & \textbf{17.76} & \textbf{0.71} & \textbf{0.58} \\
\bottomrule
\end{tabular}
\caption{Performance comparison across different stroke-count categories in QuickDraw. Classes are grouped by average stroke counts: low-stroke ($<4$), medium-stroke ($<8$), and high-stroke ($\geq 8$). Bold and underlined values indicate the best and second-best performances, respectively.}
\label{tab:performance}
\end{table*}

\begin{figure}
    \centering
    \includegraphics[width=\linewidth]{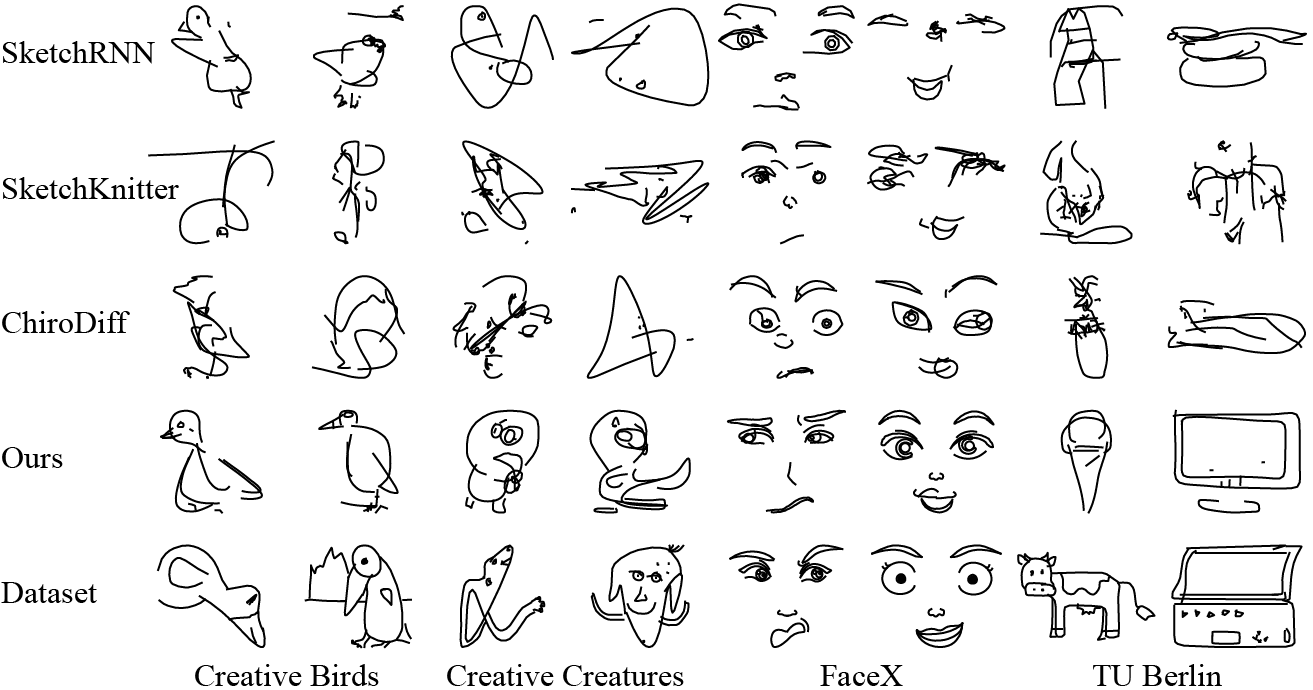}
    \caption{Qualitative generation results on several more complex datasets. Dataset names and representative examples are shown below each column for comparison.}
    \label{fig:exp-qualitative-other}
\end{figure}

\subsubsection{Datasets.}
To evaluate the model’s generalization across different sketch and face domains, we use four datasets with varying styles and complexities.

\emph{QuickDraw}~\cite{ha2017neural} contains over 50 million sketches across 345 categories. Similar to prior work~\cite{wang2023sketchknitter, das2023chirodiff}, we select 14 representative classes (e.g., airplane, face, pizza, spider) for experiments. While widely adopted in prior work, its simplicity makes it insufficient to fully demonstrate our model’s strengths, prompting us to include more challenging datasets.

\emph{TU Berlin}~\cite{eitz2012humans} consists of 20,000 sketches across 250 categories, drawn by non-artists. We use the full set to test the robustness of our method on abstract and stylistically diverse inputs.

\emph{FaceX}~\cite{cao2019ai} provides richly annotated face sketches across attributes such as gender, pose, and expression. We use only front-facing samples to evaluate spatial sensitivity in structured regions like facial features.

\emph{Creative}~\cite{ge2020creative} includes two subsets: Creative Birds and Creative Creatures, with around 10,000 sketches each. Both depict living creatures, with the former limited to birds and the latter covering a wider range of imaginative species. We train on both subsets separately to test adaptability to stylized, creative inputs.

\subsubsection{Baseline Methods.}
We compare our method with three representative vector sketch generation models: SketchRNN~\cite{yang2021sketchgnn}, SketchKnitter~\cite{wang2023sketchknitter}, and ChiroDiff~\cite{das2023chirodiff}. These three methods all focus on vector sketch generation, avoiding the limitations of traditional raster image modeling by directly outputting stroke sequences. Additionally, we include the set-based raster method, Doodleformer~\cite{bhunia2022doodleformer}, for comparison.

SketchRNN is an RNN-based variational autoencoder that can generate stroke sequences under conditional or unconditional settings and control generation diversity via a temperature parameter. SketchKnitter introduces a diffusion model framework, treating the generation of stroke points as a progressive denoising process from random initial noise to a clear sketch, significantly improving structural fidelity. ChiroDiff further incorporates a temporal modeling mechanism by resampling strokes, thereby better capturing stroke order and dynamic changes during generation. In contrast, Doodleformer is a raster-based model that predicts bounding boxes for semantic parts (e.g., head/body) and composes them into a final raster image. Its core limitations are that it does not output controllable stroke trajectories and is fundamentally restricted to datasets with semantic part annotations, such as the Creative datasets.

\subsubsection{Evaluation Metrics.}

To evaluate both the quality and diversity of generated sketches, we use three standard metrics: Fr\'echet Inception Distance (FID)~\cite{heusel2017gans}, Precision, and Recall~\cite{kynkaanniemi2019improved}. All metrics are computed over 10,000 sampled images for statistical reliability. To reduce the impact of stroke sampling resolution on feature extraction, we apply the Ramer-Douglas-Peucker (RDP) algorithm~\cite{douglas1973algorithms} for stroke simplification before evaluation.

FID measures the distributional distance between real and generated data in the Inception feature space, with lower scores indicating better visual fidelity. Precision reflects the proportion of generated samples that are close to the real data manifold (fidelity), while Recall captures how well the generated samples capture the diversity of the real data. These complementary metrics jointly assess the model's perceptual quality and coverage.

\subsection{Quantitative Comparison}

\begin{table*}[t]
\centering
\begin{tabular}{lcccccccccccc}
\toprule
\multirow{2}{*}[-0.5ex]{Method} & \multicolumn{3}{c}{Creative Birds} & \multicolumn{3}{c}{Creative Creatures} & \multicolumn{3}{c}{FaceX} & \multicolumn{3}{c}{TU Berlin} \\
\cmidrule(lr){2-4} \cmidrule(lr){5-7} \cmidrule(lr){8-10} \cmidrule(lr){11-13}
 & FID$\downarrow$ & Prec$\uparrow$ & Rec$\uparrow$ & FID$\downarrow$ & Prec$\uparrow$ & Rec$\uparrow$ & FID$\downarrow$ & Prec$\uparrow$ & Rec$\uparrow$ & FID$\downarrow$ & Prec$\uparrow$ & Rec$\uparrow$ \\
\midrule
SketchRNN & 59.85 & 0.26 & 0.28 & 121.02 & 0.44 & 0.26 & 155.02  & 0.01 & 0.31 & \underline{98.01}  & \textbf{0.73} & 0.20 \\
SketchKnitter & 59.16 & 0.25 & 0.22 & 110.46 & 0.42 & 0.27 & 156.97 & \underline{0.08} & \underline{0.34} & 99.46 & 0.56 & 0.22 \\
ChiroDiff & 60.10 & \underline{0.56} & 0.18 & 36.66 & \textbf{0.59} & 0.27 & \underline{99.33} & 0.06 & 0.30 & 98.30 & 0.53 & \underline{0.25} \\
Doodleformer & \underline{27.32} & \textbf{0.67} & \textbf{0.55} & \underline{33.46} & 0.52 & \textbf{0.69} & - & - & - & - & - & - \\
\techName\ & \textbf{26.19} & \underline{0.56} & \underline{0.30} & \textbf{19.41} & \underline{0.58} & \underline{0.32} & \textbf{7.27} & \textbf{0.76} & \textbf{0.89} & \textbf{33.68} & \underline{0.66} & \textbf{0.48} \\
\bottomrule
\end{tabular}
\caption{Performance comparison on additional datasets. Our method achieves consistently better FID, Precision, and Recall across all datasets. Bold and underlined values indicate the best and second-best performances, respectively.}
\label{tab:additional_performance}
\end{table*}

To systematically evaluate performance across varying sketch complexities, we categorize test classes into three groups by average stroke count: \textit{low-stroke} ($< 4$ strokes, including apple, moon, shoe, umbrella, and fish), \textit{medium-stroke} ($< 8$ strokes, including chair, airplane, television, face, and bus), and \textit{high-stroke} ($\geq 8$ strokes, including pizza, spider, cat, and train). Table~\ref{tab:performance} compares our method with the baselines using FID, Precision (Prec), and Recall (Rec), with top-two scores highlighted.

The results show that our method becomes increasingly advantageous as the stroke count grows. For simple sketches, where layout structure is limited, our model, designed to decouple stroke placement and shape generation, has less room to exhibit its strengths, slightly trailing ChiroDiff in some cases. However, as the stroke complexity increases and individual strokes align more with meaningful parts, our structured approach achieves clear improvements in both fidelity and consistency.

To better showcase the structural modeling ability of \techName, we also evaluate on four more challenging datasets: Creative Birds, Creative Creatures, FaceX, and TU Berlin. As shown in Table~\ref{tab:additional_performance}, our method outperforms all baselines in general, confirming strong generalization across styles and domains.

Notably, on FaceX, our gains are especially large. Facial sketches require precise spatial alignment of features like eyes and mouth, which are often distorted by sequential methods due to error accumulation. Our model avoids this by directly predicting stroke positions, leading to more stable and accurate structures.

One exception is SketchRNN's high Precision score on TU Berlin. This likely results from the dataset’s large number of categories and limited samples per class, which may cause visually plausible but semantically ambiguous outputs to match real samples in kNN-based evaluation. However, this does not reflect true structural or semantic superiority.

\begin{table*}[htbp]
\centering
\begin{tabular}{lccccccccc}
\toprule
\multirow{2}{*}[-0.5ex]{Method} & \multicolumn{3}{c}{moon} & \multicolumn{3}{c}{television} & \multicolumn{3}{c}{spider}\\
\cmidrule(lr){2-4}\cmidrule(lr){5-7}\cmidrule(lr){8-10}
 & FID$\downarrow$ & Prec$\uparrow$ & Rec$\uparrow$ & FID$\downarrow$ & Prec$\uparrow$ & Rec$\uparrow$ & FID$\downarrow$ & Prec$\uparrow$ & Rec$\uparrow$ \\
\midrule
w/o stroke normalization & 46.73 & 0.67 & 0.04 & 105.65 & 0.24 & 0.03 & 84.44 & 0.61 & 0.02 \\
w/o image-level supervision  & 38.03 & 0.57 & 0.27 & 46.52 & 0.53 & 0.13 & 139.28 & 0.11 & 0.001 \\
The full model & \textbf{25.00} & \textbf{0.68} & \textbf{0.41} & \textbf{16.28} & \textbf{0.63} & \textbf{0.55} & \textbf{15.19} & \textbf{0.71} & \textbf{0.48} \\
\bottomrule
\end{tabular}
\caption{Quantitative comparisons of stroke normalization and image-level supervision. 
Without stroke normalization, the results have degraded sketch fidelity due to inconsistent stroke representations. 
Without image-level supervision, the sketch quality and the stroke diversity are substantially compromised. 
Bold values indicate the best performance.}
\label{tab:ablation_study}
\end{table*}

\subsection{Qualitative Comparison}

\begin{figure}
\centering
\includegraphics[width=\linewidth]{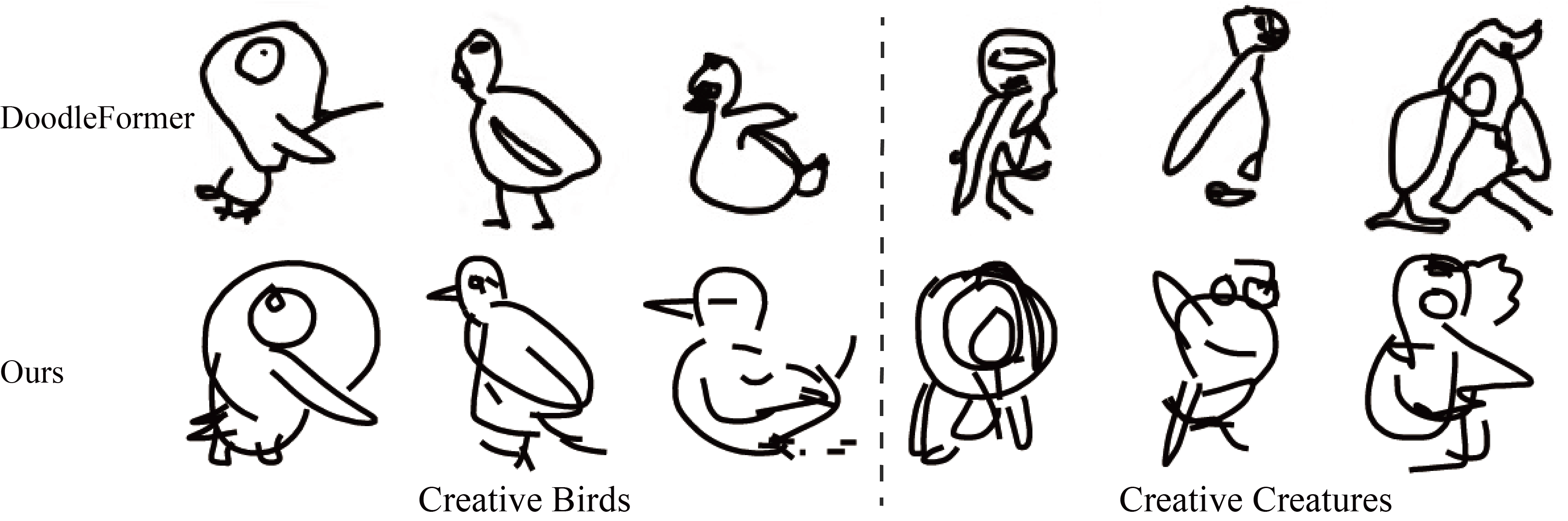}
    \caption{
    Comparison with DoodleFormer.
    For each dataset we select visually similar samples from both methods.
    }
\label{fig:exp-doodleformer}
\end{figure}

Figure~\ref{fig:exp-qualitative-quickdraw} presents the generation results of our method and the baseline methods on the QuickDraw dataset. Our approach captures a broader range of sketch variations and preserves structural diversity, while baseline methods often generate overly simplified or common shapes. In addition, our method better retains local details and demonstrates stronger consistency in global structure.

Figure~\ref{fig:exp-qualitative-other} further shows results on Creative Birds, Creative Creatures, FaceX, and TU Berlin datasets, with ground-truth examples included for reference.

For Creative Birds and Creative Creatures, the high stylistic variation and imaginative designs pose significant challenges. Diffusion-based baselines capture coarse contours but struggle with detail, while SketchRNN suffers from degraded quality in longer sequences. Our method balances structure and detail more effectively, yielding clear and recognizable outputs. We also visually compare our results with Doodleformer~\cite{bhunia2022doodleformer}, which relies on semantic part labels to generate raster sketches (see Figure~\ref{fig:exp-doodleformer}). Our approach achieves comparable visual quality while fundamentally preserving vector representations and stroke-level controllability, unlike the raster output of Doodleformer.

For FaceX, the baseline methods often produce misaligned facial features due to accumulated generation errors. In contrast, our stroke-based modeling yields more regular, semantically aligned results.

For TU Berlin, where class diversity is high and per-class data is limited, the baseline methods tend to produce either fragmented or ambiguous outputs. Our method, while still improvable, consistently produces structurally coherent and semantically interpretable sketches, showing better generalization in low-data settings.

\section{Ablation Study}

\begin{figure}
\centering
\includegraphics[width=\linewidth]{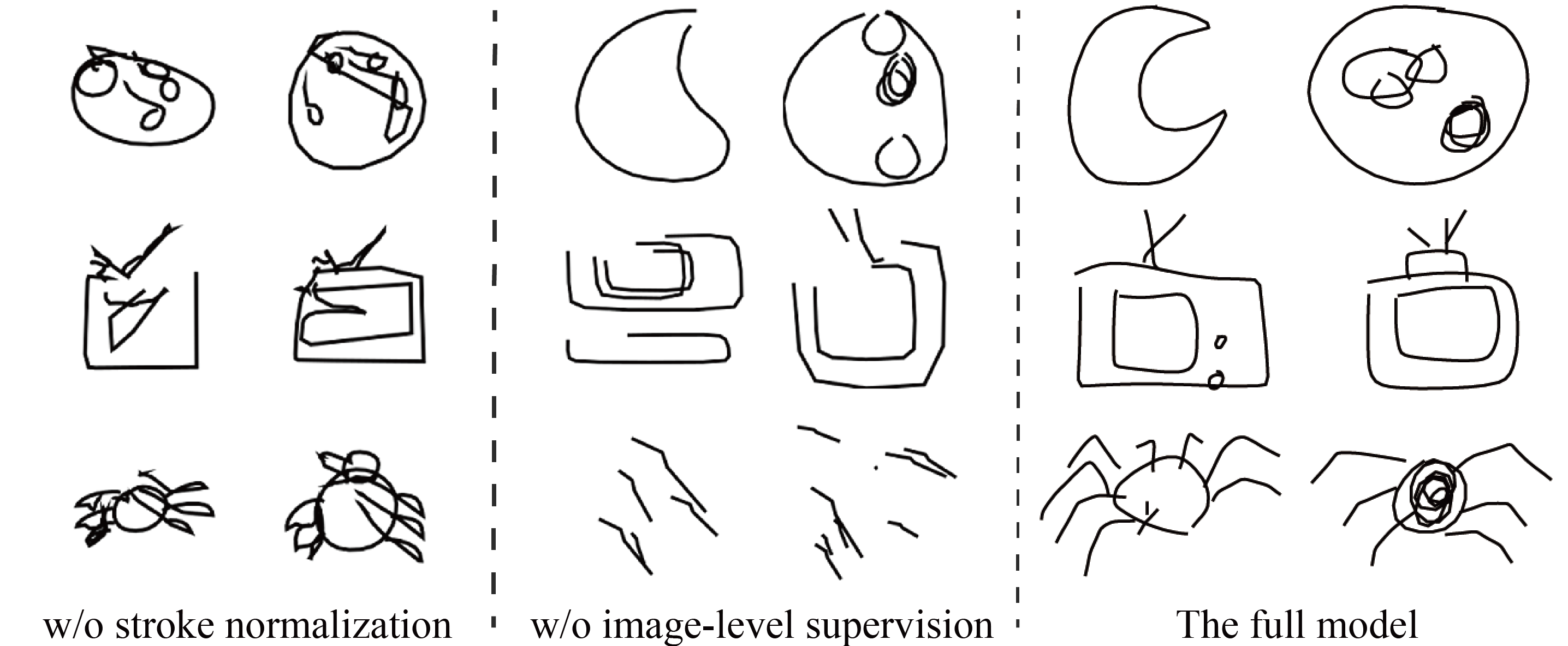}
\caption{Visualization results of the ablation study. From left to right, the columns correspond to the settings without stroke normalization, without image-level supervision, and our full model. We select three representative categories with varying stroke complexities: Moon (low-stroke), Television (medium-stroke), and Spider (high-stroke).
}
\label{fig:ab}
\end{figure}
\subsection{Stroke Normalization}

We normalize strokes by centering and scaling them into a fixed range, ensuring consistent stroke representations and reducing layout variance. To evaluate its effect, we compare with a baseline that directly uses raw strokes without normalization or explicit bounding box input.

As shown in Figure~\ref{fig:ab} and Table~\ref{tab:ablation_study}, removing normalization significantly degrades performance across both visual quality and quantitative metrics. Particularly, fine structures such as spider legs or television antennas become distorted or misplaced, indicating that normalization is essential for stable stroke encoding and structure consistency.

\subsection{Image-level Supervision}

To evaluate the impact of raster guidance, we ablate the image-level supervision by setting $\gamma=0$, effectively removing visual input.
Without image-level supervision, the model generates strokes with reduced diversity and coherence, especially in high-complexity sketches like \textit{spider}, where outputs degenerate into scattered fragments. This is further reflected in Table~\ref{tab:ablation_study}, where all metrics drop substantially. These results confirm that the visual branch plays a vital role in guiding the encoder toward learning meaningful and diverse stroke patterns.

\subsection{Limitations}  
While our method demonstrates strong performance, some limitations persist. First, complex strokes with high-frequency details remain challenging due to their inherent data intensity as point sequences, leading to suboptimal learning efficiency and generation quality for low-stroke sketches. Second, the unsigned distance field (UDF) decoding process introduces computational overhead, as we ultimately rely on point sequences for stroke representation. Future work should explore efficient implicit field decoding schemes or alternative UDF representations to streamline this component.

Additionally, since our method generates sketches with strokes as the minimal unit, it requires that the strokes in the dataset contain semantic information. Vector sketches generated by algorithms often feature strokes that span multiple semantic parts. This structure prevents our model from producing high-quality results. A potential improvement would be to segment input sketches in a more flexible, symbolically semantic manner, allowing better adaptation to such cases.

\section{Conclusion}
We have introduced \techName, a novel framework for vector sketch generation that integrates dual-modal stroke encoding with continuous diffusion modeling. Our dual-branch encoder jointly captures geometric and visual features from point sequences and stroke distance fields, fusing them into a unified latent representation. This enriched embedding space facilitates smooth feature transitions during diffusion, enabling the generation of sketches with natural stroke fluidity, structural coherence, and fine-grained detail.  Extensive qualitative and quantitative evaluations demonstrate the superiority of \techName\ in producing diverse, high-quality sketches across varying levels of complexity.

\section{Acknowledgments}
We thank the anonymous reviewers for their constructive feedback. This work was partially supported by grants from NSFC (62472287, U21B2023), Guangdong Basic and Applied Basic Research Foundation (2023A1515011297, 2023B1515120026), Shenzhen Natural Science Foundation (JCYJ20250604181519025), ICFCRT (W2441020), and the Scientific Development Funds from Shenzhen University.

%% file: Text/X_suppl.tex
\section{Data Preprocessing} 
For each sketch dataset, we begin by extracting individual strokes, which may be represented either as SVG primitives or as polylines. To construct a unified input format, each stroke is uniformly resampled into $N_p = 64$ points, which serve as model inputs. These resampled strokes are then normalized following the two-stage procedure described in the main paper, ensuring geometric consistency across sketches.

\begin{figure}[ht]
    \centering
    \includegraphics[width=\linewidth]{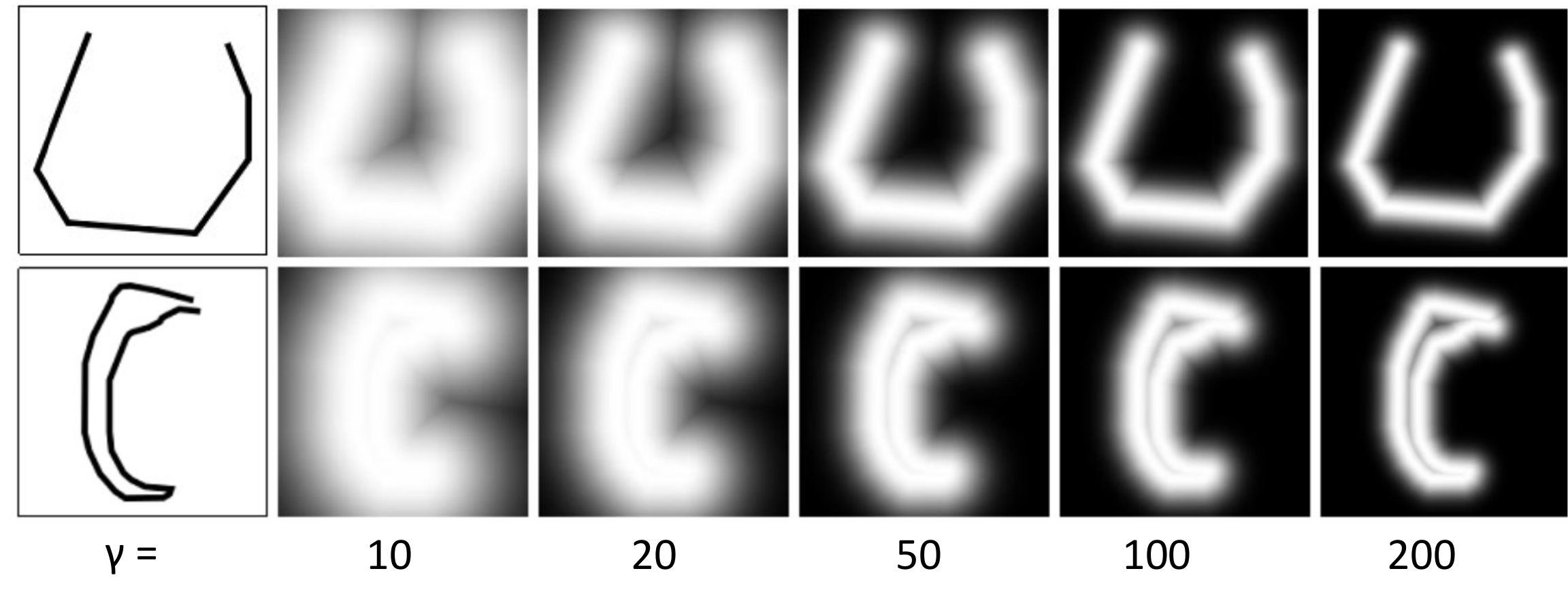}
    \caption{Unsigned distance fields generated using different values of the sharpness parameter $\gamma$. Larger values produce sharper stroke boundaries, while smaller values yield more diffuse fields. In our implementation, we set $\gamma = 50$ as a default, which provides sufficient precision without introducing artifacts.}
    \label{fig:fig_dt_gamma}
\end{figure}

To compute the unsigned distance field $I_u(g)$, we adopt the exponential decay formulation over interpolated stroke segments, using a sharpness parameter $\gamma = 50$. This avoids large empty regions produced by hard rasterization and reduces aliasing artifacts. Figure~\ref{fig:fig_dt_gamma} illustrates how different $\gamma$ values influence the resulting distance fields. Empirically, we find that model performance is largely insensitive to the exact $\gamma$ value; the key factor is whether raster visual information is incorporated at all.
In the ablation study reported in the main paper, the UDF branch is disabled by setting the decay factor to zero (removing both input and supervision), confirming that the branch contributes measurable improvements. The UDF branch adds roughly a 10

During second-stage diffusion training, we discard sketches with more than 32 strokes. For sketches containing fewer strokes, dummy strokes are padded (with visibility $v_i = -1$) to form sequences of fixed size at diffusion timestep $T$. All other per-stroke attributes (stroke embedding, bounding box) are set to zero for padded strokes. The same filtering and padding strategy is applied to all baselines for fairness.

Visibility flags are stored as continuous float values: $v_i = 0.1$ for visible strokes and $v_i = -0.1$ for absent ones. These values are treated the same as other stroke attributes in the forward and reverse diffusion processes (i.e., directly noised and denoised), without discrete Bernoulli prediction. After denoising finishes, we threshold the visibility flag at zero: values above zero indicate the stroke exists and should be decoded, while values below zero are skipped. This mechanism integrates variable-length stroke sets into the diffusion framework.

\section{Model Architecture}
Our encoder-decoder framework follows the dual-branch design described in the main paper, implemented using PyTorch Lightning. The vector encoder processes stroke point sequences of shape $(N_p, 3)$, where the third dimension includes a visibility or pen state channel. Input coordinates are first linearly projected to a hidden dimension $d_h=64$, then summed with learned positional embeddings of length $N_p=64$. Six stacked Transformer layers with 8 attention heads encode the resulting sequence. A learned query vector performs attention pooling over the encoded sequence, producing the vector representation $z_{\mathrm{seq}}$. The image encoder consists of six convolutional blocks with ReLU activations, progressively reducing the resolution of the input distance field. Feature maps are globally averaged and projected to a visual embedding $z_{\mathrm{img}}$ of dimension 64.

The fused latent vector $z_f$ is obtained by concatenating $z_{\mathrm{seq}}$ and $z_{\mathrm{img}}$, then passed through fully connected layers to predict mean and log-variance parameters for variational sampling.A KL loss is applied during the training of the encoder–decoder (before diffusion training), following standard practice in set-based latent-variable models. Although in principle the model can train without KL regularization, we observe poorer stability and more dispersed latent distributions in that case. The vector decoder takes the reparameterized latent and expands it to a fixed-length embedding sequence using a learned positional template and a 6-layer Transformer stack, producing per-point outputs $(\hat{x}_i, \hat{y}_i, \hat{m}_i)$. The image decoder mirrors the encoder structure, consisting of six transposed convolutional layers followed by a final convolution to reconstruct the distance field.

For loss computation, we adopt a weighted combination of vector reconstruction loss ($\mathcal{L}_{\mathrm{vec}}$), image loss ($\mathcal{L}_{\mathrm{img}}$), and KL divergence regularization ($\mathcal{L}_{\mathrm{KL}}$), as described in the main text. The image loss includes a perceptual term computed via a frozen VGG-16 backbone. We use $\lambda_{\mathrm{CE}}=0.1$, $\lambda_{\mathrm{L1}}=10$, $\lambda_{\mathrm{img}}=10$, and $\lambda_{\mathrm{KL}}=0.001$ as default weights.

To evaluate the impact of model capacity, we conduct ablation experiments across multiple configurations by varying the embedding size, Transformer depth, and attention width. Specifically, we explore five settings: embedding dimension $\{128, 512\}$, number of Transformer layers $\{6, 8, 16\}$, and attention heads $\{8, 16\}$, resulting in five distinct architectures. All configurations use a hidden dimension of $d_h = 64$ for input projection and a fixed maximum sequence length $N_p = 64$. Each variant is trained from scratch using the same optimization settings for fair comparison.

The final model adopts an embedding dimension of 512, with 16 Transformer layers and 16 attention heads. This configuration was selected based on a held-out validation set, where it consistently outperformed smaller models in terms of reconstruction accuracy and perceptual quality. We also applied a linear warm-up schedule of 10 epochs for this setting to stabilize training, which proved unnecessary for smaller variants. All reported results are based on this final configuration.

\section{Sketch Generation via Diffusion}
Our sketch generator is implemented as a conditional denoising diffusion model operating over latent stroke sequences. Each input is a set of latent vectors $\{\mathbf{z}_i\}_{i=1}^{N_s}$, where each vector includes a stroke embedding, bounding box parameters, and a visibility flag, as described in the main paper. Consistent with the encoder stage, all attributes (including visibility) are treated as continuous values and directly noised using the same forward process. The model learns to denoise randomly sampled sequences using a Transformer-based network without positional encodings, thereby maintaining permutation invariance across strokes.

The denoising network consists of a 16-layer Transformer with 16 attention heads per layer and an embedding dimension of 512.After the final denoising timestep, we determine the effective number of strokes by thresholding the visibility channel: strokes with predicted visibility $>0$ are decoded, while others are ignored. This integrates variable-length outputs naturally into the diffusion process. It receives the current timestep $t$, a noise-corrupted sequence, and an optional conditional embedding (reserved for conditional generation experiments). The model is trained using the DDPM framework with a linear $\beta$ schedule over 1000 timesteps. Noise is added to the input sequence according to the forward process, and the network is trained to predict the original noise using mean squared error.

After the final denoising timestep, we determine the effective number of strokes by thresholding the visibility channel: strokes with predicted visibility $>0$ are decoded, while others are ignored. This integrates variable-length outputs naturally into the diffusion process.

We adopt a learning rate of $1 \times 10^{-4}$ with the AdamW optimizer and apply a linear warm-up for the first 10 epochs. The final configuration (512-dimensional embeddings, 16 layers, 16 heads) was selected from two candidate architectures, with the alternative being a 1024-dimensional variant trained with a 100-epoch warm-up. Despite the larger model's capacity, we observed diminishing returns in validation loss and generation quality, and thus chose the more efficient 512-dimensional model for all experiments. All models were implemented in PyTorch Lightning, and the training setup allows for stable and reproducible convergence under standard hardware constraints.

\section{Training Infrastructure and Runtime}  
All experiments were conducted on a Linux server equipped with 4 NVIDIA RTX 3090 GPUs (24GB memory each). We use Python 3.11 and PyTorch 2.7.1 compiled with CUDA 12.8. To improve numerical throughput on recent GPU architectures, we enable high-precision matrix multiplication via \texttt{torch.set\_float32\_matmul\_precision('high')}.

Training the stroke encoder-decoder model takes approximately 10 hours on the 4-GPU setup, while the diffusion-based sketch generator requires around 4 hours. All experiments were implemented using PyTorch Lightning, and training is distributed using native \texttt{DDP} support to ensure scalability and reproducibility.

\section{Metric Computation}  
For quantitative evaluation, we generate 10,000 sketch samples per method and compute three widely used metrics: Fr\'echet Inception Distance (FID), Precision, and Recall. To ensure consistency, all sketches are first simplified using the Ramer-Douglas-Peucker (RDP) algorithm prior to feature extraction. All methods share the same vector-to-raster rendering pipeline before metric computation.

The FID metric quantifies the distributional discrepancy between generated and real sketches in the Inception embedding space. It is defined as:
\[
\mathrm{FID} = \|\mu_r - \mu_g\|^2 + \mathrm{Tr}\left( \Sigma_r + \Sigma_g - 2\sqrt{\Sigma_r \Sigma_g} \right),
\]
where $(\mu_r, \Sigma_r)$ and $(\mu_g, \Sigma_g)$ are the empirical mean and covariance of features from real and generated data, respectively.

Precision measures the fraction of generated samples that lie within a certain threshold $\delta$ from the real data manifold:
\[
\mathrm{Precision} = \frac{1}{N_g} \sum_{i=1}^{N_g} \mathbb{I}\left(\mathrm{dist}(g_i, \mathcal{R}) \leq \delta\right),
\]
where $g_i$ is a generated sample and $\mathcal{R}$ denotes the set of real samples.

Recall evaluates the diversity of generated outputs by computing the proportion of real samples that are closely matched by generated ones:
\[
\mathrm{Recall} = \frac{1}{N_r} \sum_{j=1}^{N_r} \mathbb{I}\left(\mathrm{dist}(r_j, \mathcal{G}) \leq \delta\right),
\]
where $r_j$ is a real sample and $\mathcal{G}$ is the set of generated samples.

The distance threshold $\delta$ is adaptively determined by computing the average $k$-nearest neighbor distance ($k=20$) among real samples in the evaluation set. This allows for a data-driven, resolution-agnostic tolerance level for all methods.This allows all methods to be compared under a data-driven, resolution-agnostic tolerance, and follows the evaluation protocol used in prior work.

\section{Additional Generation Results.}  
\begin{figure*}[t]
    \centering
    \includegraphics[width=\linewidth]{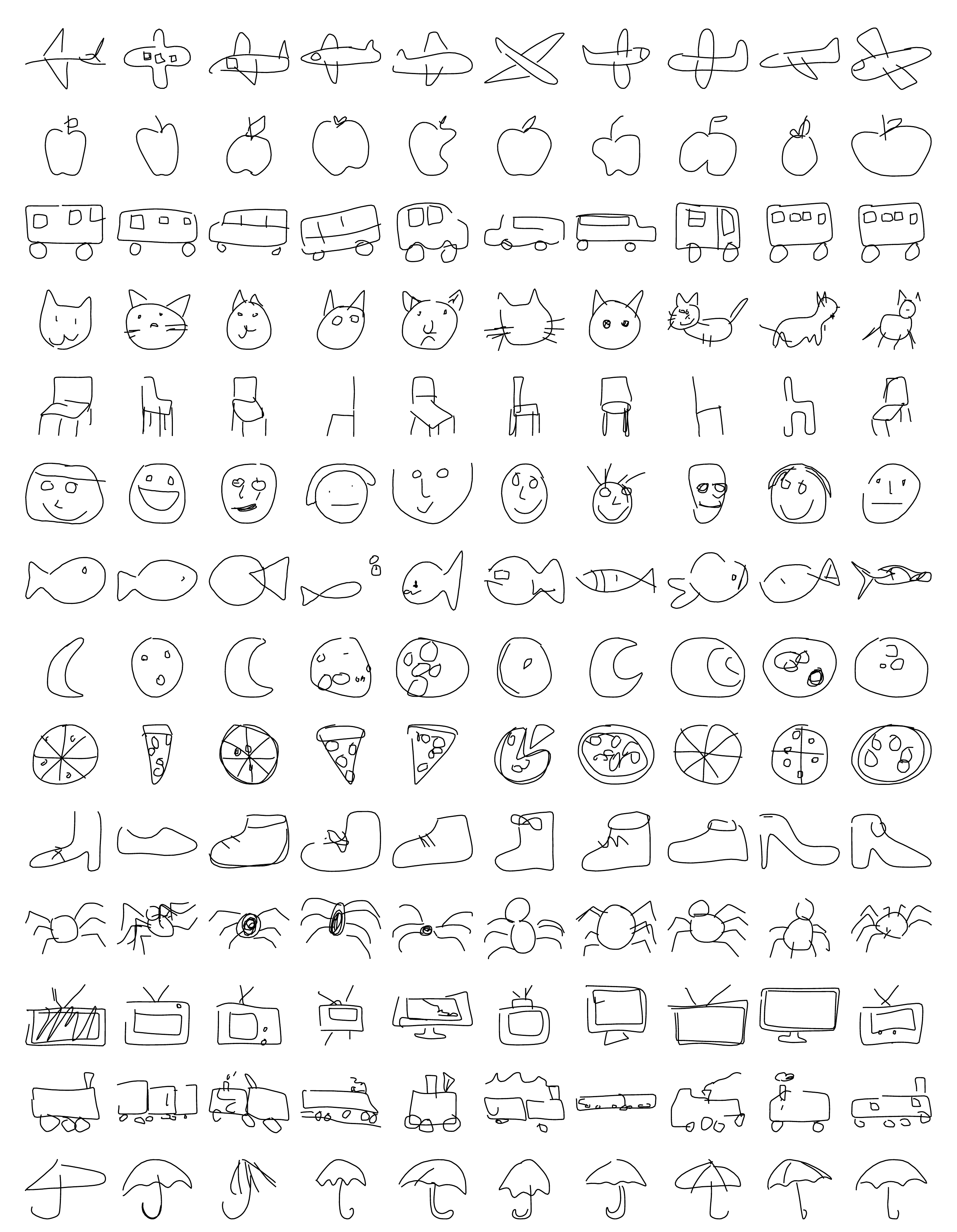}
    \caption{Additional qualitative results on 14 categories from QuickDraw. Each row corresponds to a different category: airplane, apple, bus, cat, chair, face, fish, moon, pizza, shoe, spider, television, train, and umbrella.}
    \label{fig:qd_gen_add}
\end{figure*}

\begin{figure*}[t]
    \centering
    \includegraphics[width=\linewidth]{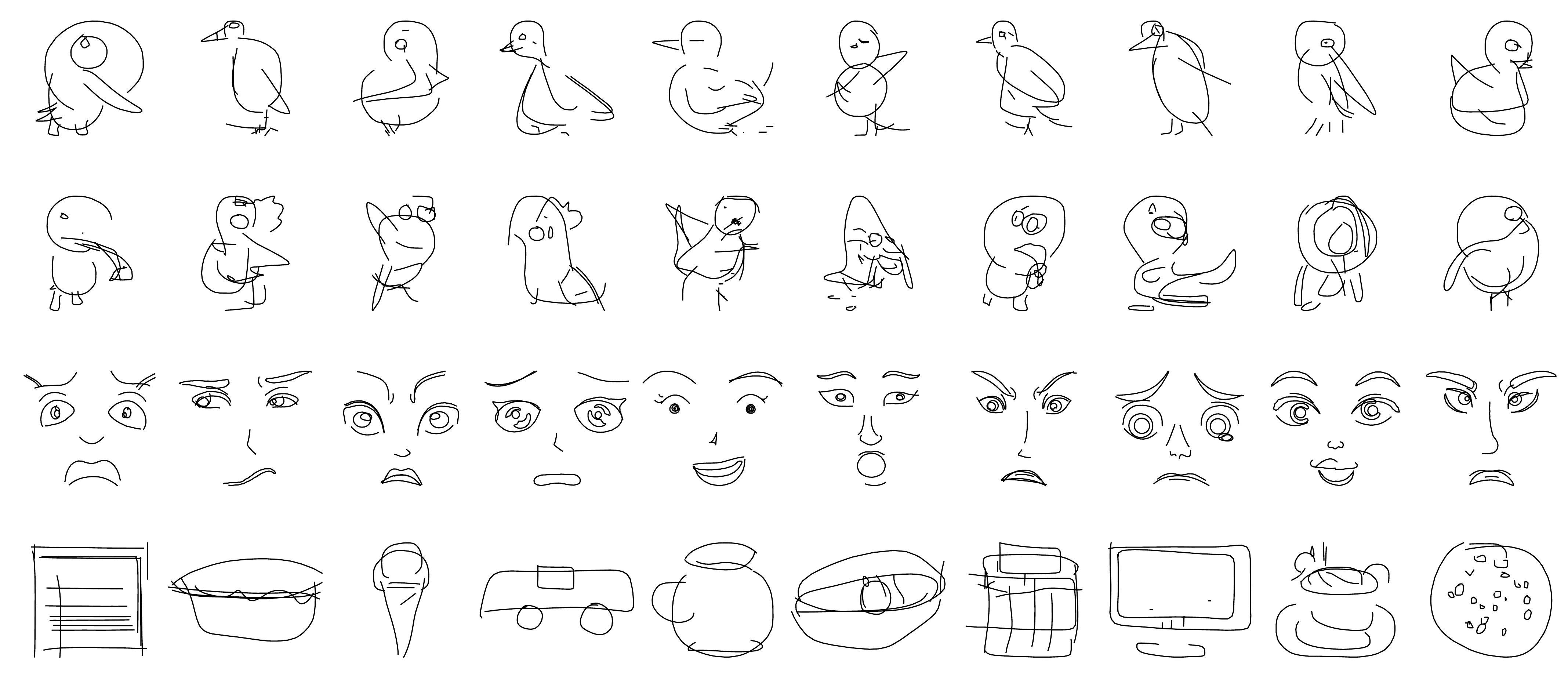}
    \caption{Additional generation results on Creative Birds, Creative Creatures, FaceX, and TU Berlin. Each row shows representative sketches from one dataset.}
    \label{fig:other_gen_add}
\end{figure*}

Figures~\ref{fig:qd_gen_add} and~\ref{fig:other_gen_add} present additional qualitative results on both QuickDraw and several more complex sketch datasets. These visualizations further demonstrate the robustness and generalizability of our model across diverse domains.

Figure~\ref{fig:qd_gen_add} showcases generated sketches from 14 different QuickDraw categories: \textit{airplane}, \textit{apple}, \textit{bus}, \textit{cat}, \textit{chair}, \textit{face}, \textit{fish}, \textit{moon}, \textit{pizza}, \textit{shoe}, \textit{spider}, \textit{television}, \textit{train}, and \textit{umbrella}, with one category per row.

Figure~\ref{fig:other_gen_add} displays representative generations on four additional datasets: Creative Birds, Creative Creatures, FaceX, and TU Berlin. Each column corresponds to a dataset, with representative examples visualized to highlight the model’s adaptability to varying styles and stroke complexities.

\begin{figure}[t]
    \centering
    \includegraphics[width=\linewidth]{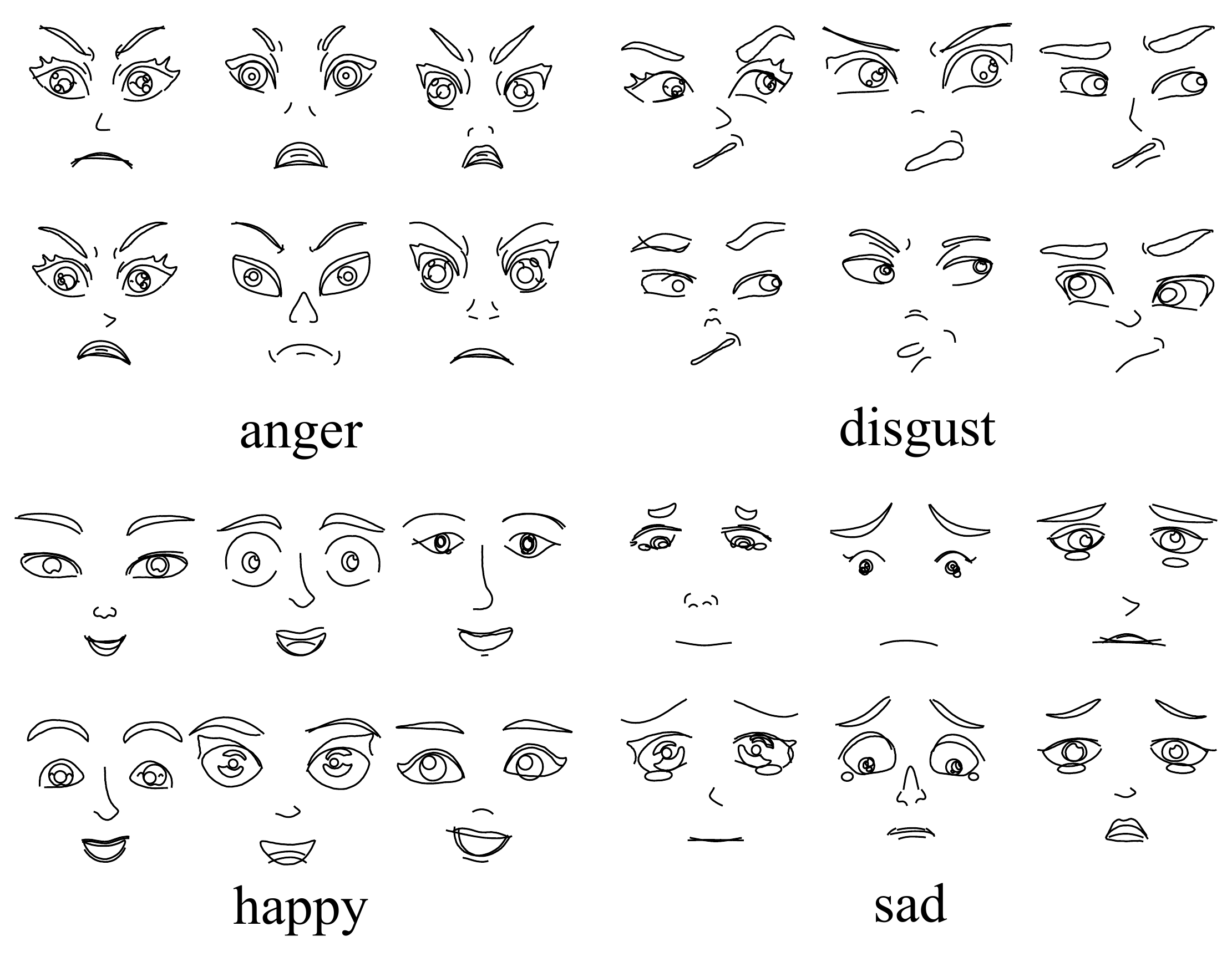}
    \caption{Class-conditional generation results on the FaceX dataset. Each block corresponds to a different emotion label: \textit{anger}, \textit{disgust}, \textit{happy}, and \textit{sad}. The model is trained once with emotion-conditioning and successfully produces coherent sketches reflecting the target emotion.}
    \label{fig:cond_facex}
\end{figure}

\begin{figure}
    \centering
    \includegraphics[width=1\linewidth]{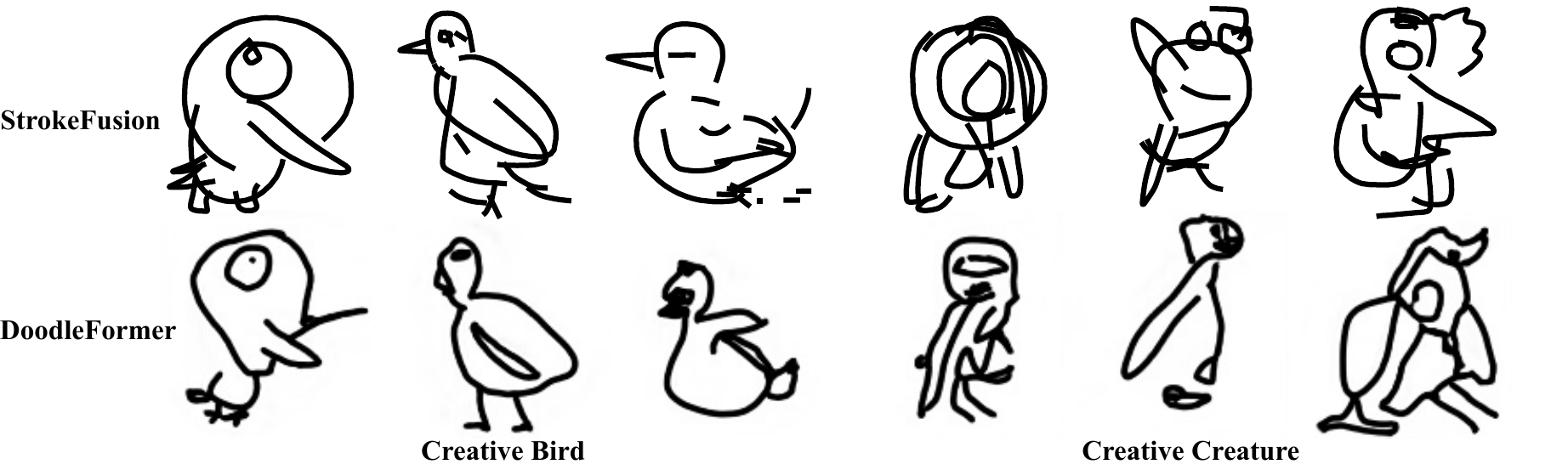}
    \caption{
    Comparison with DoodleFormer~\cite{bhunia2022doodleformer}.
    For each prompt we select visually similar samples from both methods.
    Our vector trajectories are rendered with increased stroke thickness to match the raster appearance of DoodleFormer.
    }
    \label{fig:stroke_doodleformer}
\end{figure}

\section{Conditional Generation with Emotion Labels.}  
Using the conditional embedding module described earlier, our framework supports class-conditional sketch generation. We demonstrate this capability on the FaceX dataset, where the model is trained to generate facial expressions conditioned on four emotion labels: \textit{anger}, \textit{disgust}, \textit{happy}, and \textit{sad}. Unlike training separate models for each emotion, we train a single diffusion model with emotion-aware conditioning. As shown in Figure~\ref{fig:cond_facex}, the model learns to generate emotionally consistent facial sketches with distinct and semantically aligned features across categories.

\section{Comparison with Set-based Raster Methods}
Our main paper focuses on comparing different \emph{vector} sketch generators.
Here we additionally compare our set-based vector formulation with the set-based \emph{raster} method DoodleFormer~\cite{bhunia2022doodleformer}.
DoodleFormer is trained on the Creative Birds and Creative Creatures datasets using the provided semantic part annotations (head/body): it first predicts bounding boxes for each part and then generates a raster sketch by composing part-level outputs.
Its results are only available as raster images and do not expose controllable stroke trajectories.

In contrast, our method treats each stroke as a set element without relying on explicit part labels, and directly generates executable vector trajectories.
To enable a fair visual comparison, we render our sketches with thicker strokes to approximately match the line width of DoodleFormer.
As shown in Fig.~\ref{fig:stroke_doodleformer}, our approach produces sketches of comparable visual quality while retaining vector representations and stroke-level controllability.